# Characterisation of Front-End Electronics of ChaSTE experiment onboard Chandayaan-3 lander

K. Durga Prasad*, Chandan Kumar, Sanjeev K. Mishra, P. Kalyana S. Reddy, Janmejay Kumar, Tinkal Ladiya, Arpit Patel, Anil Bhardwaj

Physical Research Laboratory, Ahmedabad, India

*durgaprasad@prl.res.in

## 1. Introduction

Chandra's Surface Thermophysical Experiment (ChaSTE) is one of the payloads flown onboard the Chandrayaan-3 lander, which is developed jointly by Physical Research Laboratory (PRL), Ahmedabad and Space Physics Laboratory (SPL) in collaboration with the various entities of VSSC, Trivandrum. The objective of the experiment is in-situ investigation of thermal behaviour of outermost 100 mm layer of the lunar surface by deploying a thermal probe. The probe consists of 10 temperature sensors (Platinum RTDs) mounted at different locations along the length of the probe to measure lunar soil temperatures as a function of depth. A heater is also mounted on the probe for thermal conductivity measurements. The onboard electronics of ChaSTE has two parts: Front-End Electronics (FEE) and processing electronics (PE). The onboard electronics of the ChaSTE payload is designed to measure lunar surface temperature with a precision of ±0.5 K in the range from 100K to 400K by using 10 Platinum RTD sensors. The front-end electronics is designed and developed by PRL, Ahmedabad, and the processing electronics is designed and developed by SPL/VSSC. The front-end electronics (FEE) card is responsible for carrying out necessary sensor signal conditioning, which includes exciting the RTD sensors, acquiring analog voltages and then converting the acquired analog signals to digital signals using an Analog to Digital Converter (ADC). The front-end card is further interfaced with the processing electronics card for digital processing and spacecraft interface. FEE needs to be calibrated and fully characterised before it is delivered for interfacing with Processing Electronics (PE) and further payload integration with the spacecraft. The characterisation details of ChaSTE FEE and its performance results are discussed in this paper.

## 2. ChaSTE Front-End Electronics

The functional block diagram of ChaSTE Front-End Electronics is shown in Figure 1. The Front-End is functionally a data acquisition-cum-signal conditioning unit. The PCB dimensions of front-end electronics measure 170mm x 100 mm. Ten four wire Platinum Resistance Temperature Detectors (RTDs) are used in the probe for temperature measurement. Each RTD channel has 4 input/output signal lines. The harness carrying all the RTD signals from the probe is connected to FEE using a 44-pin D-sub connector mounted on the chassis. The excitation current for the RTDs is generated by FEE, which is given to the RTDs through the harness. The resistance measured from each RTD is converted into voltage and is further conditioned before being sent to the multiplexer. The multiplexed signals are digitised and sent to the Processing Electronics via a 25-pin D-sub connector. The same 25-pin D-sub connector provides signals for ADC interfacing, Multiplexer operations and power. Figure 2 shows the fabricated flight (FM) version of the Front-End Card fixed into an actual ChaSTE mechanical package. Figure 3 shows the same card mounted on a zig.

We have followed a model philosophy of Engineering Model (EM), Qualification Model (QM), Flight Model (FM) and Flight Spare. Table 1 lists some of the most critical parameters of FEE to be monitored, characterised and corrected, apart from RTD sensor voltages.

## 3. Checkout Hardware and Software

Since FEE is independently developed at PRL, checkout and processing electronics is needed for its complete calibration and characterisation. A separate checkout system including processing electronics is designed based on AtMega 32A microcontroller and is used to characterise ChaSTE FEE. This processing and checkout system are interfaced to the multiplexer for RTD channel selection and in turn commanding the ADC for sending the digitised packet of analog data for 10 RTD sensors, Motor and heater currents at a sampling rate of 1s. The data is processed and averaged for 5s on-board. The electronics has provision of RS232 and USB communication with the PC for interfacing with the checkout software. In addition, processing electronics and checkout software also has wireless communication provision for remote data acquisition for laboratory and field tests. The processed data is displayed using a user-friendly software GUI developed in python. The software has provisions to establish communication using COM ports, log data for each channel at a sampling rate of 5 samples/seconds, display live voltage values for each RTD channel and also live plots of each channel individually or combined. Figure 4 shows a screengrab of Quicklook and Checkout Software GUI. The screengrab shows the temperature reading in multiple channels as a function of time.

## 4. Calibration of the Front-End Electronics of ChaSTE

Figure 5 shows the temperature measurement scheme used in FEE for one RTD channel [1]. The key characterisation parameters of FEE are shown in Figure 5 and Table 1.

The offsets for each Front-End Electronics (FEE) card is mainly characterised by the parameters in Table 1 which are first corrected. Long-term channel stability and temperature variation levels after stability corrections are key parameters indicating the performance of the FEE card.

*Table 1: Characteristics of ChaSTE FEE$^\$$*

| Parameter | Name | Remarks |
|---|---|---|
| $V_{CC}$ | Positive Voltage Bias | These are external parameters, measured at the connector pins. * |
| $V_{EE}$ | Negative Voltage Bias | |
| $I_{CC}$ | Positive Bias Current | |
| $I_{EE}$ | Negative Bias Current | |
| $\Delta V_{ch}$ | Channel Voltage Variation | Measured for each RTD channel# |
| $\Delta T_{ch}$ | Channel Temperature Variation | |

*\* These are measured values for external operating parameters. Any deviations in these parameters will be monitored and corrected first before going ahead for calibration and characterisation.*

*# Deviations in these parameters are due to the deviation in the component values form the ideal ones. This will provide a unique set of values for each FEE card.*

$V_{CC}$, $V_{EE}$, $I_{CC}$, $I_{EE}$ are voltages and currents fed to each RTD channel in the FEE. These are the currents and voltages being measured at the respective pins of the connector. These are basically the current and voltages from an external programmable power supply provided to the card. While the voltage fluctuations from the power supply are very well taken care by the LDOs, any small

variation needs to be accounted for. $\Delta V_{ch}$ is the variation in the voltage levels of the channel outputs at a given constant temperature. $\Delta T_{ch}$ is the variation in temperature of the channel output at a given constant temperature. $I_{exc}$ is the excitation current given to each RTD as per the design requirements. A predefined and fixed constant excitation current is required for the measurement of the RTD resistances and any small variation also needs to be accounted and corrected for. The subsequent section discusses the variations in individual parameters and their effect on final temperature measurement. The deviations (offsets) in the parameters listed in Table 1 are unique values for each FEE card dictated by the input voltages/currents and deviations in the values of the electronic components from their ideal values. These deviations (offesets) are first measured and corrected, further which the FEE is finally calibrated and characterized for each model of the card. This entire calibration and characterisation process are divided into two parts. First is the Electronic Calibration and the second is the Temperature Calibration.

Electronic Calibration aims to find the voltage offsets for all the RTD channels for a particular FEE card which remains constant. This combined offset in the output voltage is mainly due to the deviations in the values of the electronic components used with respect to their designated values. For the electronics card, the RTDs are essentially resistive loads. For the PT1000 RTDs, the limit resistance values (w.r.t. corresponding temperatures) likely to be encountered at the landing site are listed in Table 2. This table gives an idea of the lower limit, higher limit and typical lunar day time temperatures during the entire operation of ChaSTE. The electronics design has a margin on both sides of the limit values and the calibration covers these design margins right from Liquid nitrogen temperatures of ~77K to ~400K. Since the RTD response is almost linear throughout this range, the electronics offset at any of the specified resistive loads will be the actual offset. For this purpose, we have prepared calibration resistive loads of 1 kΩ, 1.2 kΩ, 330 Ω and so on, using very low tolerance (0.1%) SMD resistors. These resistive loads termed as 'dongles' can be plugged in at the probe harness connector directly to simulate particular temperatures. We have used a 1 kΩ (273 K) dongle for offset correction and have used the 330 Ω and 1.2 kΩ dongles for validating the offsets. The detailed procedure is given in the protocol sub-section.

*Table 2: PT1000 Resistance and Temperature likely to be encountered at Chandrayaan-3 landing site*

| Temperature (K) | PT1000 Resistance (Ω) |
|---|---|
| 403 K | 1500 Ω |
| 325 K | 1200 Ω |
| 273 K | 1000 Ω |
| 197 K | 700 Ω |
| 122 K | 330 Ω |

Once electronics offsets are corrected, the conversion from voltage to temperature is done. After voltage offset correction of the FEE card, the temperatures also need to be calibrated with respect to a standard thermal reference. For this purpose, we have used a thermo-well (Fluke 9190A Field Metrology Well) along with certified reference thermometer of resolution 0.001 K to calibrate the temperatures. This thermo-well is capable of maintaining temperatures from 178 K to 383 K at a resolution of 0.01 K. The probe and Thermometer are inserted in the thermo-well using a custom made insert fabricated in-house at PRL. The detailed procedure of temperature calibration is given in the protocol sub-section.

**4.1 Calibration Protocol**

After completion of wiring of the card, basic functional test (FT) is carried out with the card mounted on a zig after which the card is transferred from the zig and fitted into the ChaSTE chassis.

All further tests, calibration and characterisation activities are done with the FEE Card mounted in ChaSTE chassis.

For calibration, the package is kept in a laminar flow test bench in a clean room (Class-1000) at PRL. All the electrical/signal connection of the front-end as well checkout card are made and the appropriate power levels for front-end and checkout card is set using a programmable power supplies. A logbook is maintained to record all the activities, parameters and values for traceability and verification.

The protocol followed to carry out the calibration is as follows:
1. First, a 1K dongle is connected. The corresponding voltage response of each channel is logged for correction of the electronic offset within each channel. These offsets are further incorporated in the checkout software for automatic offset correction.
2. Next, the other dongles (1.2 kΩ and 330Ω) are connected to verify the offset corrections made earlier.
3. Next, an identical probe, fabricated especially for this activity, mimicking the functional requirements of ChaSTE Probe, is placed in a thermo-well referred earlier, for temperature calibration of the FEE.
4. Temperature range from ~178K(-95°C) to 383K(110°C) is covered with appropriate temperature intervals of 10K. At each temperature, the data is logged for about 10-30 minutes for channel stability verification.
5. All the data is analysed to arrive at the characterisation of parameter values listed in Table 1. The data logged using the thermo-well gives the temperature stability at each temperature for a wide range of temperatures.

Figure 6 shows the setup for the calibration of the FEE of ChaSTE FM card. Similar setup was used for the calibration of the QM and FM Spare Cards as well. Figure 7 shows the calibration curve for sensors 7,8 and 9 of the FM card. For comparison, a linear fit is also shown along with the calibration curves.

**4.2 Possible deviation from ideal conditions**

Considering the percentage deviations for values of the passive and active components used in the FEE design, we obtain the percentage error deviation in the temperature ($\Delta T$) as follows,
$$\Delta T = \Delta R_f + \Delta R_i + \Delta V_{ref} + \Delta R + \Delta R_{RTD}$$
Where,

$$\Delta V = \Delta G + \Delta V_i$$
$$\Delta G = \Delta R_f + \Delta R_i$$
$$\Delta V_i = \Delta I_{exc} + \Delta R_{RTD}$$
$$\Delta I_{exc} = \Delta V_{ref} + \Delta R$$

The channel voltage $V$ is a function of the resistance of the RTD and is given by:
$$V = G * V_i$$

Where $V_i$ is the voltage output of the previous stage and $G$ is the gain. The gain is same for each channel, and is constant over the entire measurement range. The gain ($G$) is further given by
$$G = \frac{R_f}{R_i}$$
Where $R_f$ and $R_i$ are the feedback and non-inverting input resistance of the gain stage OpAmp. The voltage ($V_i$) for each RTD is obtained using following equation:

$$V_i = I_{exc} \times R_{RTD}$$

where, $I_{exc} = \dfrac{V_{ref}}{R}$

Table 3 lists the values of the primary (independent) parameters and also their variations

*Table 3: FEE sensor and primary electrical parameters and their expected tolerances*

| Parameter | Value | Tolerance |
|---|---|---|
| $V_{ref}$ | 2.5 V | ±0.12% |
| $R$ | 8.25 kΩ | ±1 % |
| $R_{RTD}$ | variable | ±0.12% |
| $R_f$ | 2.1 kΩ | 0.1 % |
| $R_i$ | 390 Ω | 2 % |

Table 4 lists the acceptable limits and observed variations for secondary (or derived) electrical parameters. A very important point to be noted here is that the tolerance is a measure of the temperature dependent variation in the value from its ideal value. For a given temperature, as was the case with the electronics package that was being calibrated, the error due to temperature change was almost zero. Apart from the temperature-dependent deviations, any other error in the values of the parameters that affects the final measurement of the RTD channel voltage are reflected in the channel offset voltage. These offsets are constant at a constant temperature and are corrected for. On top of this, any variation in the RTD channel voltage is seen in the stability plot. As per the design of the FE electronics, a voltage variation of ±4.15 mV corresponds to a temperature variation of ± 0.5 K. Therefore, as long as the stability of the RTD channel output is within ±4.15 mV at a constant temperature, the accuracy of temperature measurements will be better than ± 0.5 K.

*Table 4: Secondary (derived) electrical parameters and their variations*

| Parameter | Acceptable Limit | Measured Variation |
|---|---|---|
| $\Delta V_{ch}$ | ± 4.15 mV | ± 1 mV |
| $\Delta T_{ch}$ | ± 0.5 K | ± 0.12 K |

he possible deviations in all these primary and secondary parameters are listed in tables 3 and 4. It may be noted that the percentage variations or tolerances given in Table 3 is for the whole measurement range and maximum possible errors. At a given temperature, the deviations are calculated using temperature drifts in resistance values and the drift in the voltage value of $V_{ref}$. Table 4 shows the combined effect of all possible variations and their acceptable limits. As evident from the Table, the temperature accuracy obtained from the calibration and subsequent testing activities is well within the acceptable limits. The offset plot, stability plot and other relevant results are discuesed in detail in the results section.

## 5. Characterisation ad Test Results
### 5.1. Functional Test Procedure and Thermal FT

The Front-End Electronics Card has been fabricated as per the fabrication flow chart, standard protocols and other QA/QC requirements. The card also underwent the process of conformal coating and Local Potting. Functional tests (FT) were carried out at different stages of FEE fabrication, especially after completion of a stage of wiring or the card being subjected to any mechanical, thermal or electrical process, to assess the effect of the process on the performance of the card. One such representative process described here was the thermal soak test of the FEE Card. For this, an Initial Functional Test (IFT) was conducted before the thermal soak and Final Functional Tests (FFT) were conducted after the completion of the thermal soak test to evaluate the electronics performance of the card. The thermal soak cycle is shown in Figure. 8.

*Table 5: Average channel output values during Pre and Post FT of ChaSTE FEE during thermal soak test*

| Channel No. | IFT (V) | FFT (V) | ΔV (mV) |
|---|---|---|---|
| Channel 1 | 1.648 | 1.646 | -2 |
| Channel 2 | 1.645 | 1.645 | 0 |
| Channel 3 | 1.648 | 1.645 | -3 |
| Channel 4 | 1.647 | 1.645 | -2 |
| Channel 5 | 1.642 | 1.64 | -2 |
| Channel 6 | 1.648 | 1.647 | -1 |
| Channel 7 | 1.653 | 1.650 | -3 |
| Channel 8 | 1.642 | 1.641 | -1 |
| Channel 9 | 1.647 | 1.645 | -2 |
| Channel 10 | 1.647 | 1.646 | -1 |

After thermal soak test, functional verification tests were carried out. This was done using a custom-made setup in an ESD safe workbench placed in a Class 1000 clean room of PRL. All the functional tests were carried out for a standard input temperature of 273 K (by using 1 kΩ resistance dongle as probe input) for all channels. The output of each channel is monitored for its stability for a minimum of 10 minutes. Considering the 1 LSB ADC error and other electronics related offset/deviations, the acceptable level of stability is nearly ±4.15 mV (corresponding to nearly ±0.5 K) within the duration of the test, which was the requirement of the functional verification test.

### 5.1.2 Pre-Thermal FT

The test setup is shown in Figure 9. The Front-End card was tested in lab conditions with lab power supplies for 10 RTD channels. Initially, an Engineering Model was connected to check for power levels and readings were logged for 10 minutes post which the Flight Model was connected. A constant 1 kΩ resistance dongle is connected to the probe input connector of the FM FE card. The data was logged for stability/offset readings for two hours under ambient conditions.

### 5.1.3 Post-Thermal FT

Active Thermal Soak Test for 2 hrs cold and 2 hrs hot were carried out at suggested ChaSTE payload temperature levels as per ISRO ETLS document for Chandrayaan-3. The soak temperature and their duration is as shown in Figure 8. The PRL thermal chamber facility was used for this thermal testing with proper ESD and grounding facilities available at the work bench. The package was kept inside the chamber with thermistors placed on the package and the base plate for temperature monitoring. Post-Thermal Final FT was carried out with the package kept under ambient lab conditions as per the procedure described earlier. The setup used for thermal soak test is shown in Figure 10 and the thermal cycle is shown in Figure 8. The output voltages obtained from ChaSTE FE for all temperature channels during pre- and post-thermal soak test are shown in Table 5. Observations from this test show that the deviation under all conditions durong thermal soak test was < ±3 mV (considering all influencing factors), a value well within the acceptable limits of design, and thus establishing a good functional performance of ChaSTE FEE card. Similar performance is observed during all other functional tests as well.

### 5.2. Characterisation of the Front-End Electronics of ChaSTE

After the electronic calibration and functional tests, the characterisation of the electronics using actual ChaSTE probe and under simulated lunar environment is carried out. This is the final step to ascertain the behaviour of the electronics when subjected to the actual probe in simulated lunar conditions. To this end, experiments were conducted in the in-house built lunar simulation chamber[2] and the data is processed to characterise the response of the FEE.

To achieve this at PRL, we made use of a Lunar Simulation Chamber at SIMPEX lab [2], [3] which is capable of simulating lunar surface strata, temperatures and pressures. The chamber is shown in Figure 11. This chamber is completely in-house designed and fabricated and is capable of achieving vacuum of the level of ~$10^{-7}$ torr. Also, the chamber can accommodate soil strata with provisions for heating and cooling to attain temperature range from around Liquid Nitrogen Temperature (77 K) to 430 K. The chamber has multiple feedthroughs and has a large view port for electrical interfaces and optical view.

The characterisation process involves 3 steps as described below.

1. The lunar analog samples are prepared and a stratum is made in a sample cup filled up to a height such that all the sensors of the probe are immersed inside the sample with the top one being just beneath or on the surface.
2. The ChaSTE Probe is placed in the centre of the sample cup and the external systems are connected to the probe and other internally present devices via the feedthroughs.
3. The Front-End Card along with the Checkout were operated, the data was logged and processed.

A logbook is maintained that logs all the information, observations, status of the systems part of the characterisation setup. Information such as the Mass, Volume and Bulk Density of the sample is measured and recorded. The vacuum level inside the chamber, the chamber temperature, the ambient temperature etc. are also recorded at defined intervals of time. All this data is helpful in analysing and interpreting the data generated during the characterisation process. At the very beginning, after powering on the front-end electronics and connecting either the dongle or the probe, measurement of electrical parameters mentioned in Table 1 are made.

The processing of the data obtained from the characterisation experiment is carried out. The offset curve is plotted and offset corrections to individual channels are made. Offset plot for FM is shown in Figure 12. Channel stability, which defines the variation in temperature over an extended period of time when sensors are kept at a constant temperature, is also plotted and the maximum variation w.r.t. the actual values are calculated. Figure 13 shows the stability plot for FM of ChaSTE FEE. It can be clearly seen from figures 12 and 13 that the variability in temperatures of all channels are around ±1 mV (i.e. ±0.12 K), a value well within our acceptable limits and hence establishing the functional performance of ChaSTE FEE.

The Calibration, Characterisation and Functional Testing of the Front-End Electronics of ChaSTE was done as explained in the previous sections. The test and calibration process of FM yielded the results as compiled in the Table 4.

$\Delta V_{ch}$ is obtained by calculating the maximum upper and lower variation in output voltage of each channel during the entire period of observation. Similarly, $\Delta T_{ch}$ is obtained by calculating the maximum upper and lower variation in derived temperatures of each channel during the entire period of observation. $V_{CC}$, $V_{EE}$, $I_{CC}$, $I_{EE}$ is measured at corresponding pins of the connector.

It has been observed that the deviations in measured and expected values of the parameters are well within the range of permissible limits for each parameter under all conditions. Figure 12 shows the plot obtained for individual RTD channels at ~273 K (1 KΩ dongle). It can be seen from figure 12 that all the RTD channels have some offset with respect to the expected value of around ~1.710 V (as per the design). This is due to the deviation of the component values of each RTD channels

(from their expected value) which will be a unique set of values for this FM card. These are the offsets/deviations which are first corrected before proceeding for further calibration and characterisation. Figure 13 shows the channel stability plot after offset correction for two different channels 4 and 9, at around 265 K (-8°C) temperature. Figure 13 clearly demonstrates the channel stability of around 1 mV which is well within the acceptable limit of the design. Similar trend is evident for other channels as well. Figure 14 shows the heating curve under simulated lunar conditions with the ChaSTE probe placed in a two-layer stratum of lunar analogous soil, to mimic the actual experimental conditions. As seen in figure 14, the temperature evolution at different depths of the soil sample is clearly recorded by temperature sensors of the ChaSTE probe with a good temperature stability thus proving the credibility of the experiment.

## 6. Summary and Conclusion

The calibration, characterisation and functional test activities of Front-End Electronics of ChaSTE were carried out with the objective of testing and ensuring proper functioning, electronics offset correction and characterising the FEE in simulated lunar environment. A 2-phase calibration process involving electronic offset correction and temperature calibration was carried out followed by a thermal soak and functional testing in between the thermal cycle. All these activities were successfully completed and the results from them provided us with a really good understanding of the behaviour of the FEE under different thermal and electrical conditions as well as when subjected to the simulated conditions of the actual ChaSTE experiment. The performance of the card was very much within the design margins and its behaviour in simulated lunar environment was as desired. The data from these activities is useful in the interpretation of the actual science data of ChaSTE.

## Acknowledgements

We acknowledge the anonymous reviewer for his useful comments and suggestions which has greatly improved the presentation and content of the manuscript.

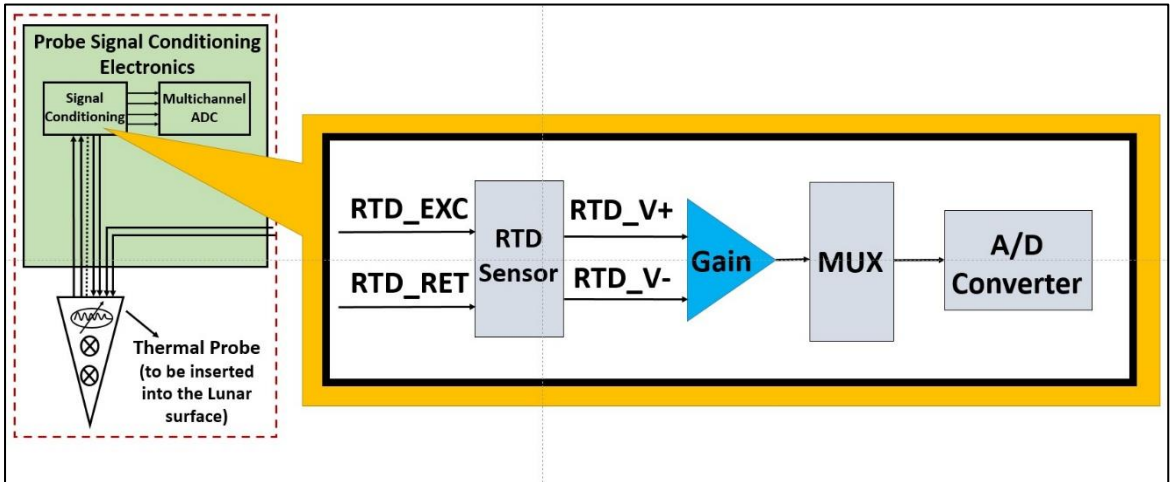

*Figure 1 Functional Block Diagram of Front-End Electronics of ChaSTE*

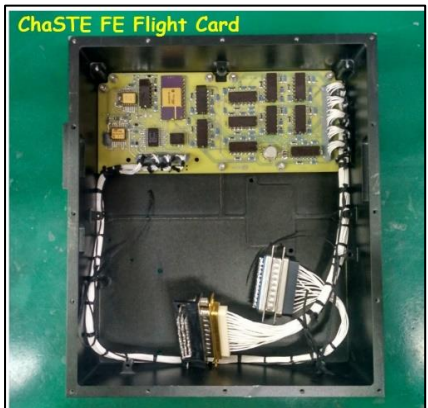
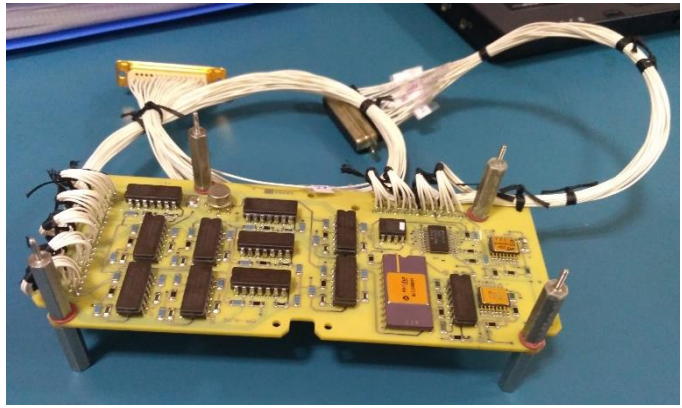

*Figure 2: FM Card in ChaSTE Box*     *Figure 3: ChaSTE FEE FM Card mounted on an aluminum zig*

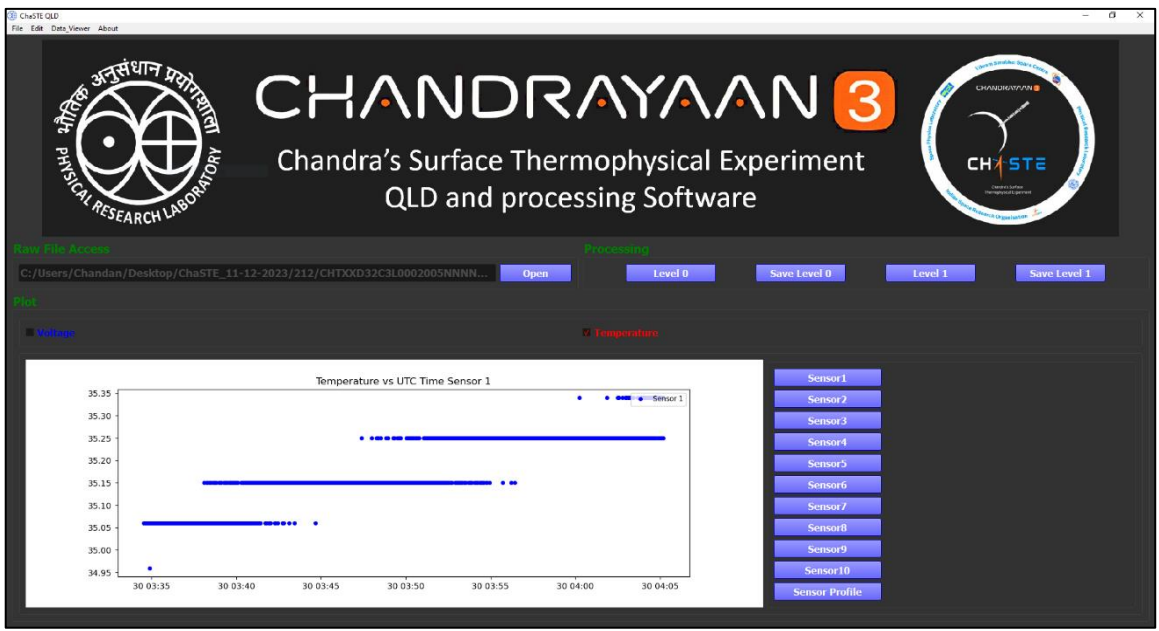

*Figure 4: A screengrab of the checkout Software GUI for calibration, characterization and functional testing of ChaSTE*

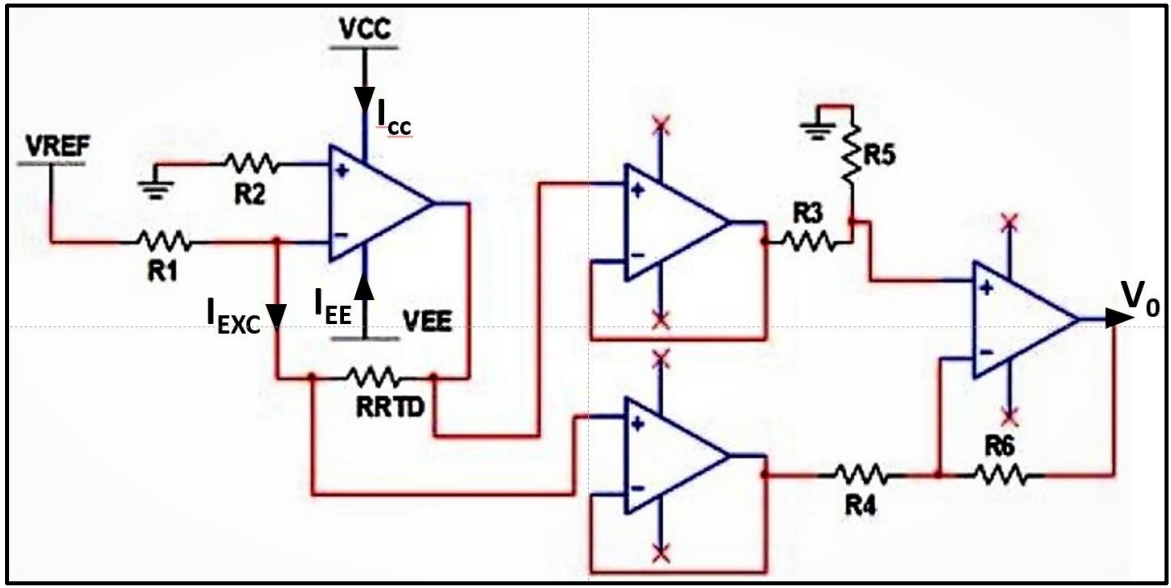

*Figure 5: Measurement scheme for a single RTD channel*

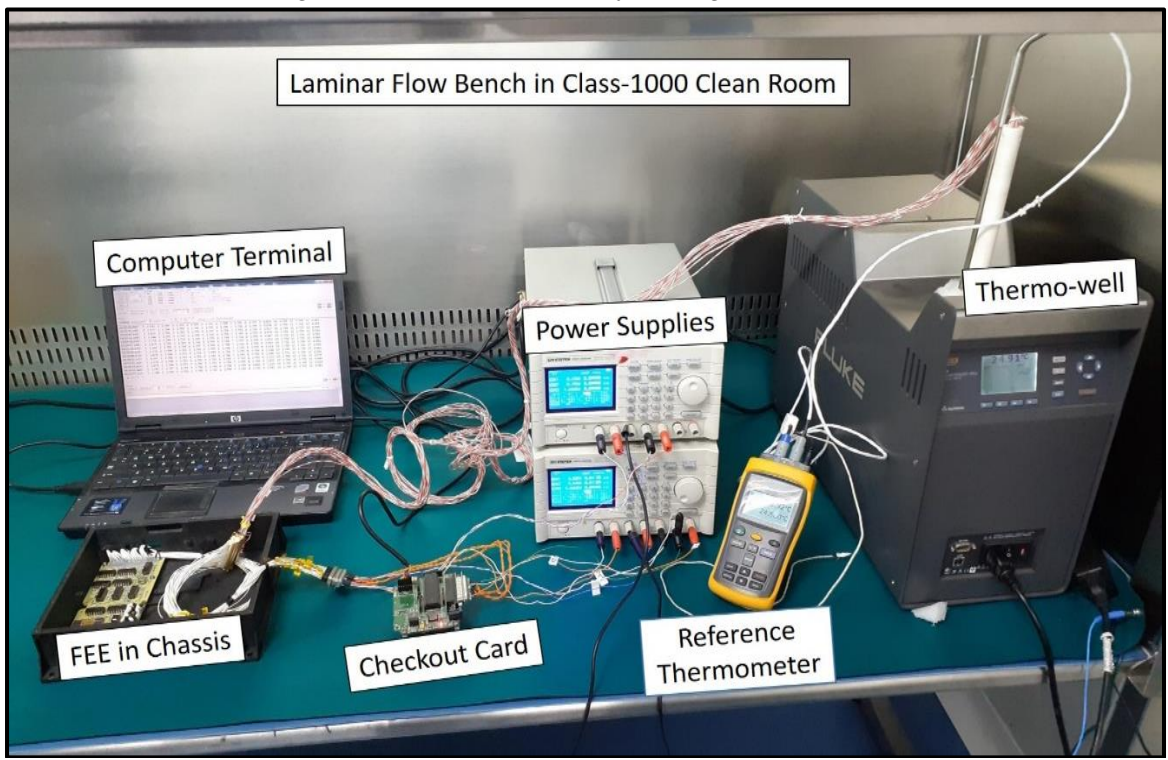

*Figure 6: Test setup for calibration of ChaSTE QM Card*

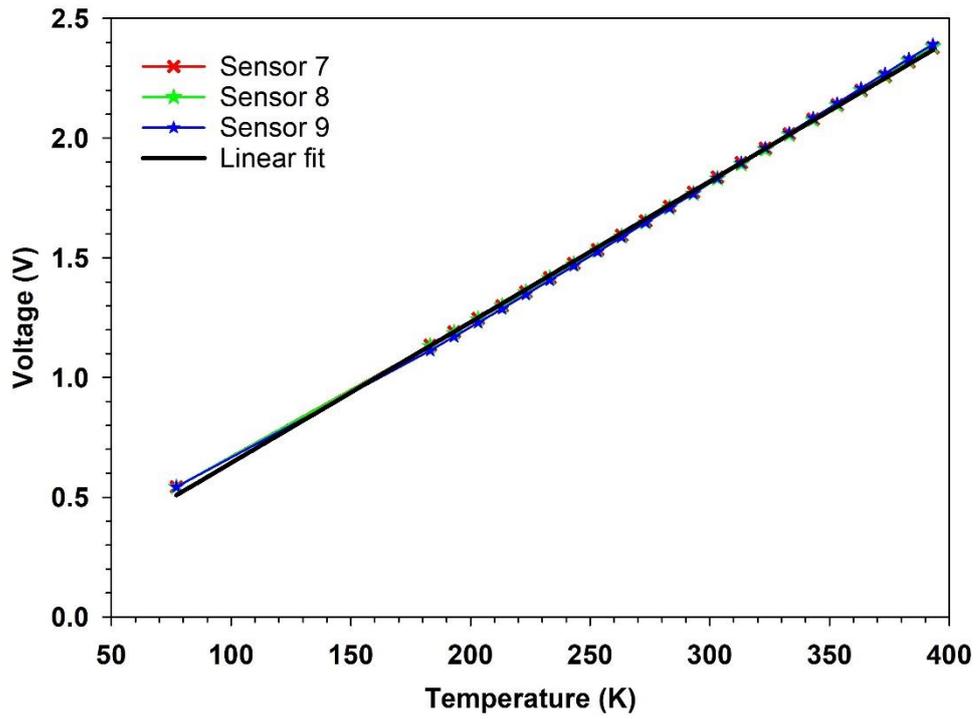

*Figure 7: Calibration curve for a few sensors of ChaSTE FE FM Card*

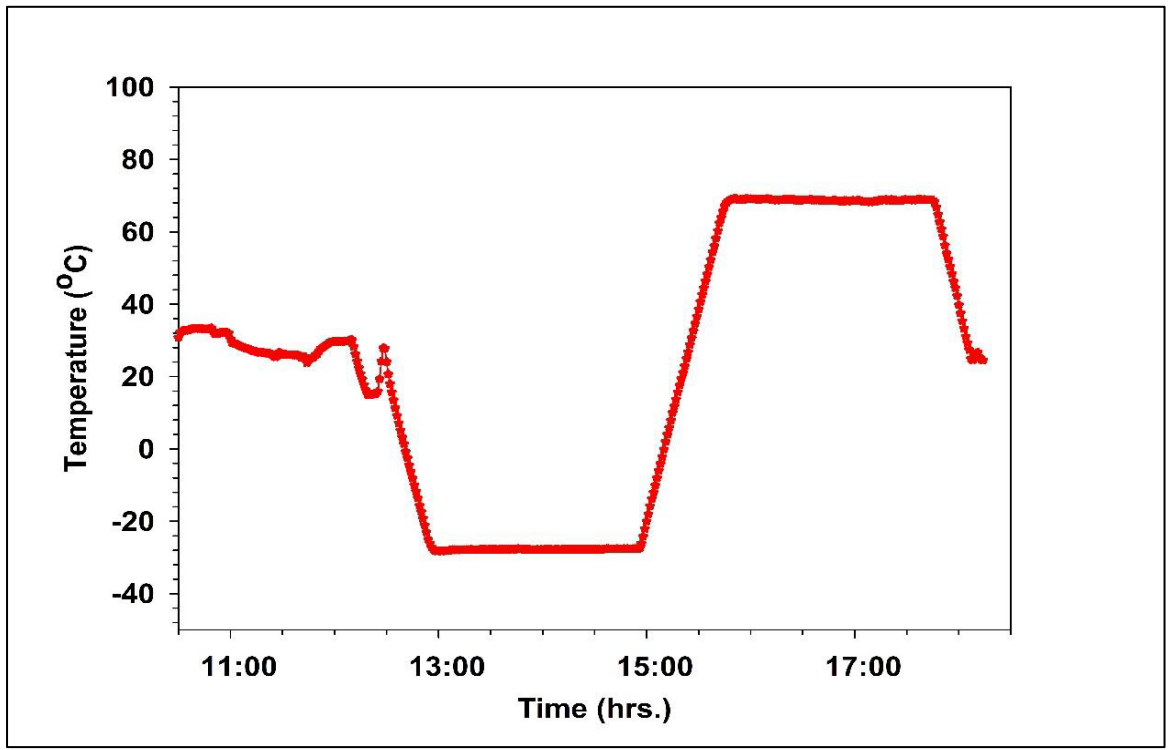

*Figure 8: Thermal Cycle used for Cold and Hot Soak*

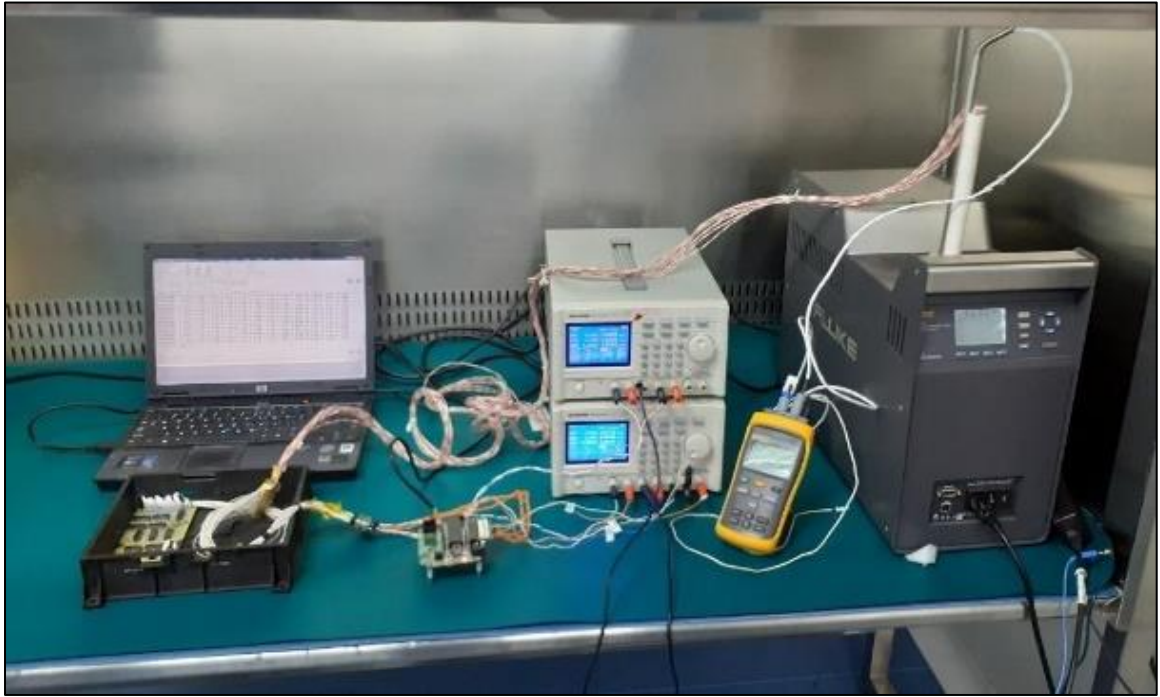

*Figure 9 Functional Testing Setup for ChaSTE Front-End Electronics*

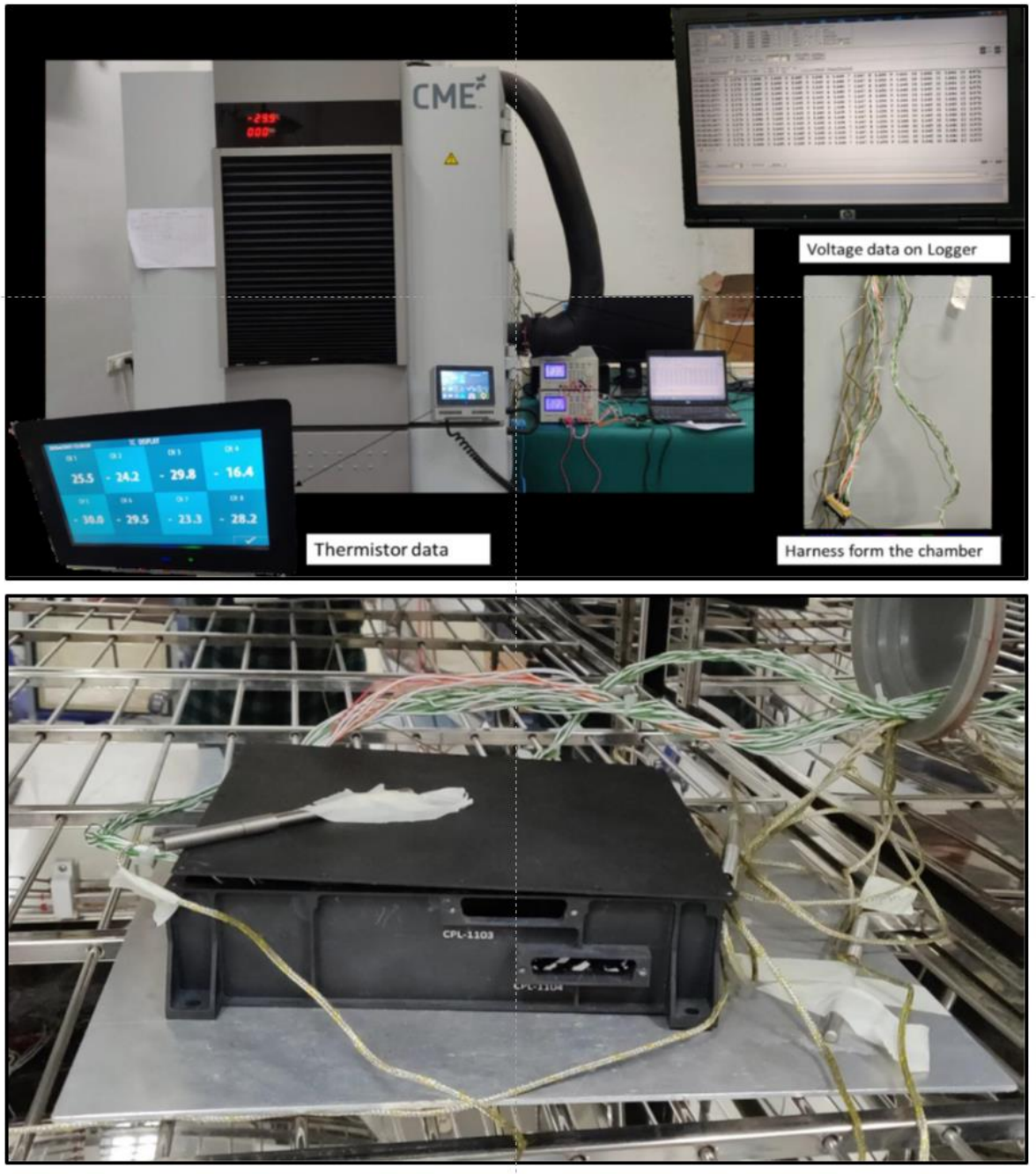

*Figure 10: Thermal Soak Test Setup used for ChaSTE FE Testing. Complete Test-Setup (Top); Package placed inside the thermal chamber (bottom)*

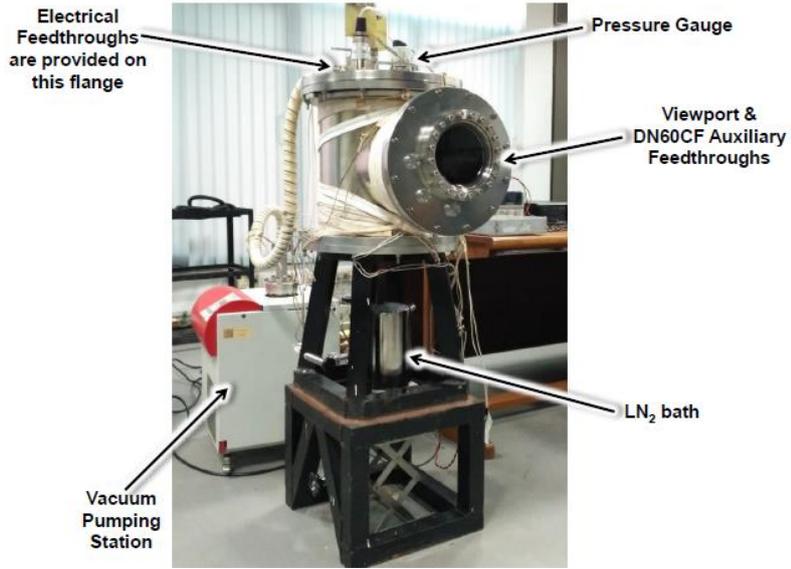

*Figure 11: In-House Built Lunar Simulation Chamber[2]*

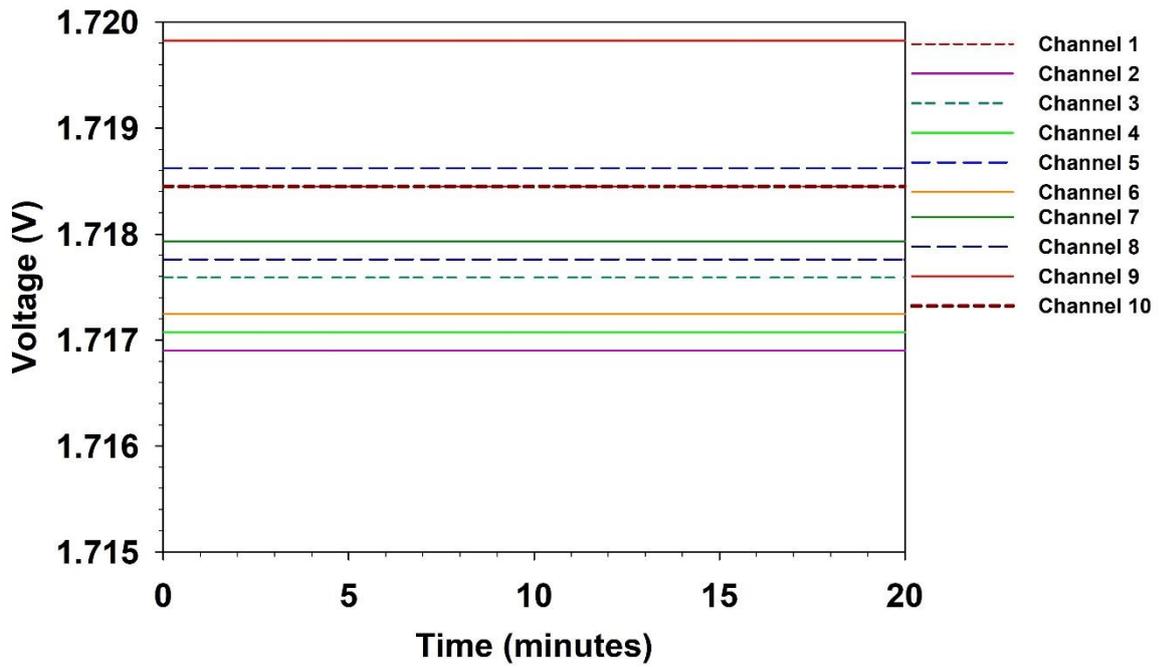

*Figure 12: Voltage offset curve for ChaSTE FE QM Card*

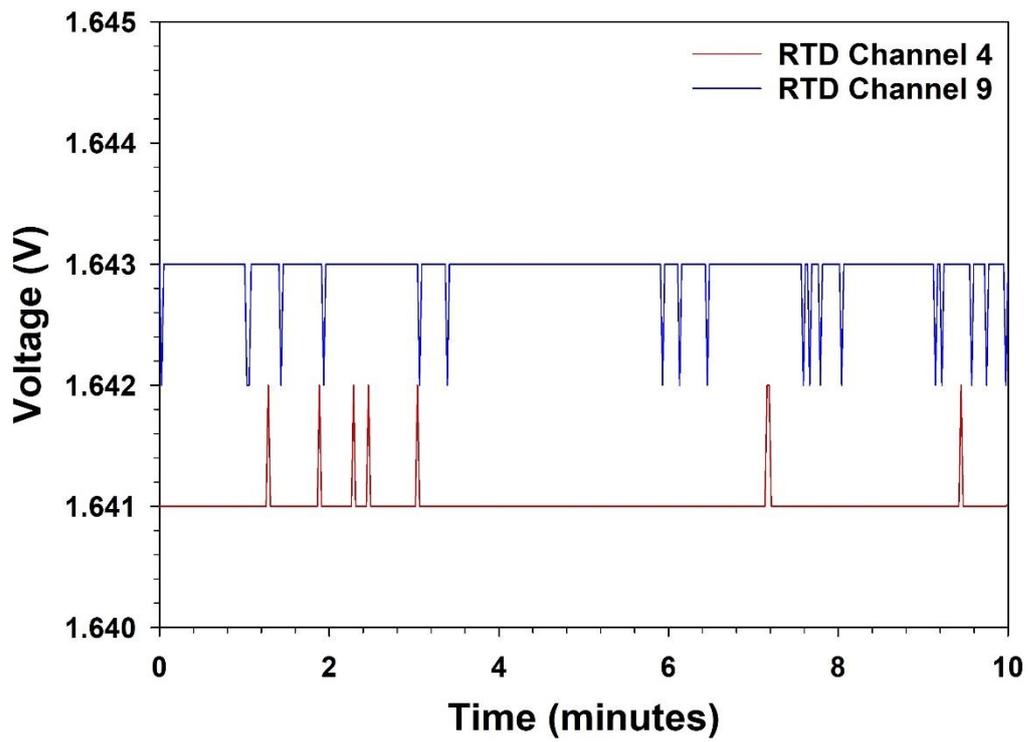

*Figure 13: Voltage stability plot for ChaSTE FE FM card*

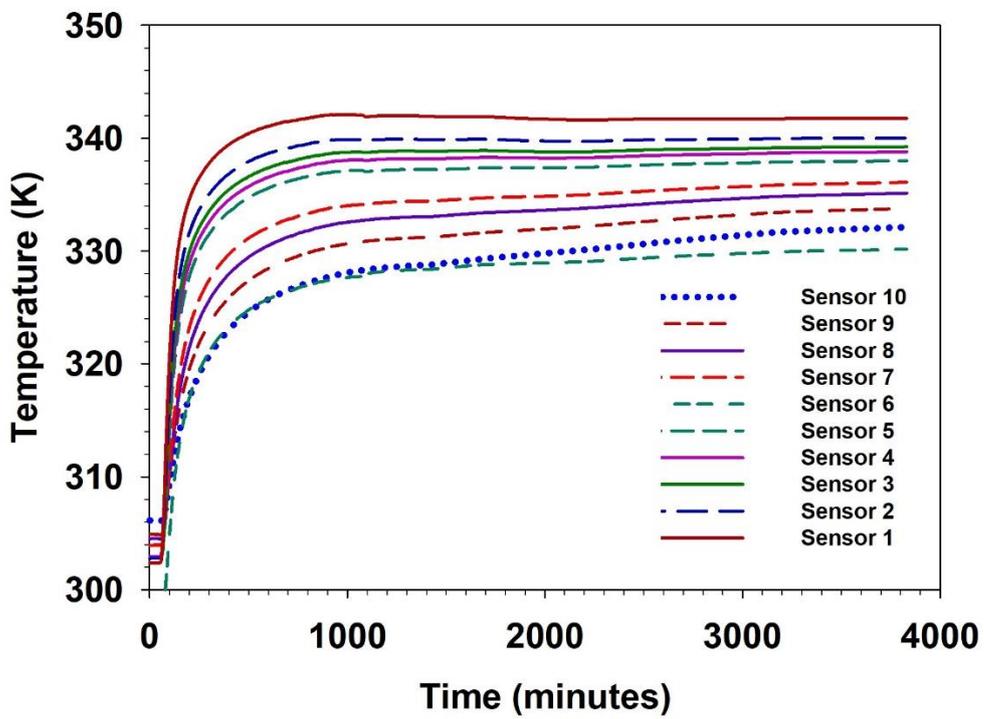

*Figure 14: The heating curve with ChaSTE probe inserted in a strata made of Lunar analog samples in Lunar Simulation Chamber*